\documentclass[fleqn,usenatbib]{mnras}

\usepackage{newtxtext,newtxmath}

\usepackage[T1]{fontenc}
\usepackage{ae,aecompl}

\usepackage{graphicx}	
\usepackage{amsmath}	
\usepackage{amssymb}	

\hypersetup{draft}

\title[SALT/SAAO observations of AT 2017gfo/GW170817]{A comparison between SALT/SAAO observations and kilonova models for AT 2017gfo: the first electromagnetic counterpart of a gravitational wave transient $-$ GW170817
}

\author[D. A. H. Buckley et al.]{David A. H. Buckley,$^{1,2}$\thanks{E-mail: dibnob@saao.ac.za (DAHB)}
Igor Andreoni$^{3,4,5}$, Sudhanshu Barway,$^{1}$, Jeff Cooke$^{3,4,6}$,  
\newauthor
Steven M. Crawford$^{1,2}$, Evgeny Gorbovskoy$^{7}$, Mariusz Gromadski$^{8}$, Vladimir Lipunov$^{9,7}$,
\newauthor
Jirong Mao$^{10,11,12}$, Stephen B. Potter$^{1}$, Magaretha L. Pretorius$^{13,1}$, Tyler A. Pritchard$^{3}$,
\newauthor
Encarni Romero-Colmenero$^{1,2}$,
Michael M. Shara$^{14,15}$, Petri V\"ais\"anen$^{1,2}$
\newauthor
and Ted B. Williams$^{1}$ 
\\
$^{1}$South African Astronomical Observatory, P.O. Box 9, Observatory 7935, Cape Town, South Africa\\
$^{2}$Southern African Large Telescope Foundation, P.O. Box 9, Observatory 7935, Cape Town, South Africa\\
$^{3}$Centre for Astrophysics and Supercomputing, Swinburne University of Technology, PO Box 218, H29, Hawthorn, VIC 3122, Australia\\
$^{4}$The Australian Research Council Centre of Excellence for Gravitational Wave Discovery (OzGrav), Australia\\
$^{5}$Australian Astronomical Observatory, 105 Delhi Rd, North Ryde NSW 2113, Australia\\
$^{6}$The Australian Research Council Centre of Excellence for All-Sky Astrophysics (CAASTRO), Australia\\
$^{7}$M.V.Lomonosov Moscow State University, Sternberg Astronomical Institute, Universitetsky pr., 13, Moscow, 119234, Russia\\
$^{8}$Warsaw University Astronomical Observatory, Al. Ujazdowskie 4, PL-00-478, Warszawa, Poland\\
$^{9}$M.V.Lomonosov Moscow State University, Physics Department, Leninskie gory, GSP-1, Moscow, 119991, Russia\\
$^{10}$Yunnan Observatories, Chinese Academy of Sciences, 650011 Kunming, Yunnan Province, China\\
$^{11}$Center for Astronomical Mega-Science, Chinese Academy of Sciences, 20A Datun Road, Chaoyang District, 100012 Beijing, China\\
$^{12}$Key Laboratory for the Structure and Evolution of Celestial Objects, Chinese Academy of Sciences, 650011 Kunming, China\\
$^{13}$Department of Astronomy, University of Cape Town, Private Bag X3, Rondebosch 7701, South Africa\\
$^{14}$Department of Astrophysics, American Museum of Natural History, Central Park West and 79th Street, New York, NY 10024, USA\\
$^{15}$Institute of Astronomy, University of Cambridge, Madingley Road, Cambridge CB3 0HA, UK\\
\\
}

\date{Accepted XXX. Received YYY; in original form ZZZ}

\pubyear{2017}

\begin{document}
\label{firstpage}
\pagerange{\pageref{firstpage}--\pageref{lastpage}}
\maketitle

\begin{abstract}
We report on SALT low resolution optical spectroscopy and optical/IR photometry undertaken with other SAAO telescopes (MASTER-SAAO and IRSF) of the kilonova AT 2017gfo (aka SSS17a) in the galaxy NGC4993 during the first 10 days of discovery. This event has been identified as the first ever electromagnetic counterpart of a gravitational wave event, namely GW170817, which was detected by the LIGO and Virgo gravitational wave observatories. The event is likely due to a merger of two neutron stars, resulting in a kilonova explosion. SALT was the third telescope to obtain spectroscopy of AT 2017gfo and the first spectrum, 1.2 d after the merger, is quite blue and shows some broad features, but no identifiable spectral lines and becomes redder over time. We compare the spectral and photometric evolution with recent kilonova simulations and conclude that they are in qualitative agreement for post-merger wind models with proton: nucleon ratios of $Y_e$ = 0.25 $-$ 0.30. The blue colour of the first spectrum is consistent with the lower opacity of the Lathanide-free r-process elements in the ejecta. Differences between the models and observations are likely due to the choice of system parameters combined with the absence of atomic data for more elements in the ejecta models. 
\end{abstract}




\section{Introduction}
Following the advanced LIGO detection of the gravitational wave transient, GW170817/G298048 \citep{LIGO2017, Essick2017,Connaughton2017}, and its simultaneous detection as a short gamma ray burst by the Fermi GBM \citep{von Kienlin2017}, the optical counterpart was subsequently identified by \citet{Coulter2017} as a point source, located ~10 arcsec from the center of the S0 galaxy, NGC4993, initially named SSS17a and then renamed AT 2017gfo, following the IAU naming convention.

The source was independently identified by several groups following the refinement of the error position provided by the LIGO/Virgo G298048 BAYSTAR HLV map \citep{Singer2017}. The optical transient of GW170817 was subsequently identified to be a kilonova \citep{Kasliwal2017, McCully2017, Nicholl2017}, the remnant of a neutron star $-$ neutron star (hereafter abbreviated as NS) merger \citep{Blinnikov1984}. The radioactive decay of r-process elements in the expanding wind or envelope of a kilonova has been postulated to explain the energetics plus spectral and photometric evolution \citep[e.g.][]{Li1998, Rosswog1999, Rosswog2005, Freiburghaus1999, Metzger2010,  Barnes2016, Coughlin2017, Kasen2017, Tanaka2013, Tanaka2017}.

In this paper we show the SALT spectra of the kilonova, AT 2017gfo, taken respectively at 1.2 and 2.2 d after the GW trigger, and compare them to the three different models developed by \citet{Tanaka2017} for kilonova ejecta. These models have varying degrees of opacity and abundances of the lanthanide r-process elements. The flux and the blue nature of the first SALT spectrum seems to be most consistent with the ''blue" kilonova wind model, with a proton/nucleon ratio of $Y_e$ = 0.30, for the assumed distance of 40 Mpc. We also present photometry of AT 2017gfo over a period 10 days post detection, in optical ($B,~V~\&~R$) and infrared ($J,~H~\&~K$), derived from the MASTER-SAAO and IRSF telescopes at Sutherland, respectively. We again compare these results to the respective kilonova model predictions, and conclude that either the $Y_e$ = 0.25 or 0.30 wind models are qualitatively similar to the observed magnitudes.

\section{SALT and SAAO Observations}
Following the detection of AT 2017gfo, the optical counterpart to GW170817 \citep{Coulter2017, Lipunov2017}, director's discretionary time observations (programme 2017-1-DDT-009) were undertaken on 18 \& 19 August 2017 on the Southern African Large Telescope (SALT; \citet{Buckley2006}). The observations were taken with the prime focus Robert Stobie spectrograph \citep{Burgh2003}, beginning in twilight and proceeding until the end of the available telescope track time. A third attempt on 20 August resulted in no meaningful data being obtained due to the sky brightness coupled with the degree of fading of the kilonova. The observational details are included in Table 1, and preliminary reports on the results are presented in \citet{Shara2017, Abbott2017, McCully2017, Andreoni2017}.

The low resolution PG300 surface relief transmission grating was used, rotated to an angle of 5.75$^{\circ}$, with a long slit of width 2'', which implies a $\sim$88\% slit throughput in the given seeing conditions. The spectra had a resolution which varied from R $\sim$ 150 (at $\sim$ 3750\AA) to $\sim$ 400 (at $\sim$ 9600\AA), with a mean of  R $\sim$ 380. The observations were reduced using the PySALT package \citep{Crawford2010}, which accounts for basic CCD characteristics (cross-talk, bias and gain correction) and removal of cosmic ray cleaning, wavelength calibration, and relative flux calibration. Additional reductions to account for accurate sky and galaxy background removal were done using standard IRAF routines.

Because of the SALT design, which has a moving, field-dependent and under-filled entrance pupil, absolute flux calibration with SALT is difficult to achieve with a good degree of accuracy, which at best is $\pm20\%$. Observations were taken in morning twilight on 18 Aug 2017 of the spectrophotometric standard star EG21, which was used to determine a relative flux calibration on both nights. The spectral fluxes were then corrected by convolving the observed spectra with standard Johnson-Cousins $B~\&~R$ filters and comparing the results with $B~ \&~R$ observations taken simultaneously with MASTER-SAAO facility, which are presented in \citet{Lipunov2017} and are included in Table 1. This comparison implied that the spectra were required to be adjusted in flux by a multiplicative constant of 2.04 and 2.4, respectively, on the two nights. MASTER-SAAO also observed in a filter-less mode on several nights, defined as $W$ (see Table 1), which is between the $B$ and $V$ filters, depending on the object colour \citep[e.g.][]{Lipunov2010}. AT 2017gfo was also observed by the SAAO 1.0 m Elizabeth telescope, however the data quality was too poor to estimate meaningful magnitudes.

Details of the infrared observations of AT 2017gfo using the Infrared Survey Facility (IRSF) and the data reduction are described in \citet{Kasliwal2017}.  The near-infrared ($J, H~ \&~K_s$) simultaneous imaging camera, SIRIUS, installed on the 1.4 m telescope IRSF (InfraRed Survey Facility) telescope was used in the period 23, 24 and 26 August 2017, up to 9.2 days after the GW trigger time.  A total of 10  dithered exposures of 30 sec each with dithering radius of 60 arcsec per observing sequence, respectively, were observed and repeated typically seven to eight times to obtain good S/N ratio. Dark frames were obtained at the end of the nights and twilight flat frames were obtained before and after the observations. The data reduction includes dark frame subtraction, flat-field correction, sky-subtraction, dither combination and astrometric calibration, and was carried out using the SIRIUS data reduction pipeline software.

\section{Comparison of observed spectra to kilonova models}
Following the merger of two neutron stars, the cause of the GW170817 event \citep[e.g.][]{LIGO2017}, material is ejected in a kilonova explosion (also referred to as macronova), whose luminosity is powered by the radioactive decay of r-process nuclei \citep{Kasen2017, Tanaka2017}. Here we compare our two SALT spectra to the recently derived dynamical ejection models and high $Y_e$ ($Y_e$ = proton: nucleon ratio) models of post-merger ejecta by \citet{Tanaka2017}, for delay times of 1.5 and 3.5 days following a kilonova explosion. We have determined the predicted flux densities using the model luminosities together with the assumed distance to NGC 4933/AT 2017gfo of 40 $\pm$ 8 Mpc \citep[e.g.][]{Abbott2017}.  It would appear that the dynamical ejecta model, APR4-1215 \citep{Hotokezaka2013, Tanaka2013, Tanaka2017}, for two merging NSs of 1.2 and 1.5 $M_{\odot}$ and ejecta mass of $M_{ej}$ = 0.01 $M_{\odot}$ is both too red and too under-luminous compared to the observations (see the red curve in Fig. 1). The lower velocity post-merger ejecta model, with $Y_e$ = 0.3, qualitatively matches the observed flux for $\lambda$ $<$ 5200\AA, while there is a deficit of flux at longer wavelengths. 

The second SALT spectrum, taken 2.2 d following the GW event, is shown in Fig. 2, together with the same respective models used previously, but for 3.5 d post-merger (these were the next oldest models after t = 1.5 d). These models are a poorer match to the data, although they are $\sim$ 1.3 d older than the observations. Since the models for t = 1.5 d are closer in time to the observations ($\sim$ 0.7 d younger), we show the $Y_e$ = 0.25, t = 1.5 d model as well, which is a closer match to the observation both in flux and colour.  

\section{Comparison of observed magnitudes to kilonova models}
We undertook a similar comparison between the kilonova models of \citet{Tanaka2017} and the optical-IR photometry of AT 2017gfo, taken at the SAAO during the first $\sim$10 days of the kilonova outburst (Fig.3.), assuming a 40 Mpc distance.

In general it appears that the models are somewhat under-luminous in comparison with the observations. While the $Y_e$ = 0.30 model (Fig. 3) is in better agreement with the observed magnitudes, particularly in the optical region, the $Y_e$ = 0.25 model (Fig. 4) seems to show an overall better agreement if the model was $\sim$ 1.5 magnitudes brighter. These discrepancies are likely due to different values for key  parameters (e.g. masses of the NSs and ejecta) and missing elements in the models.

\begin{figure}
	\includegraphics[width=\columnwidth]{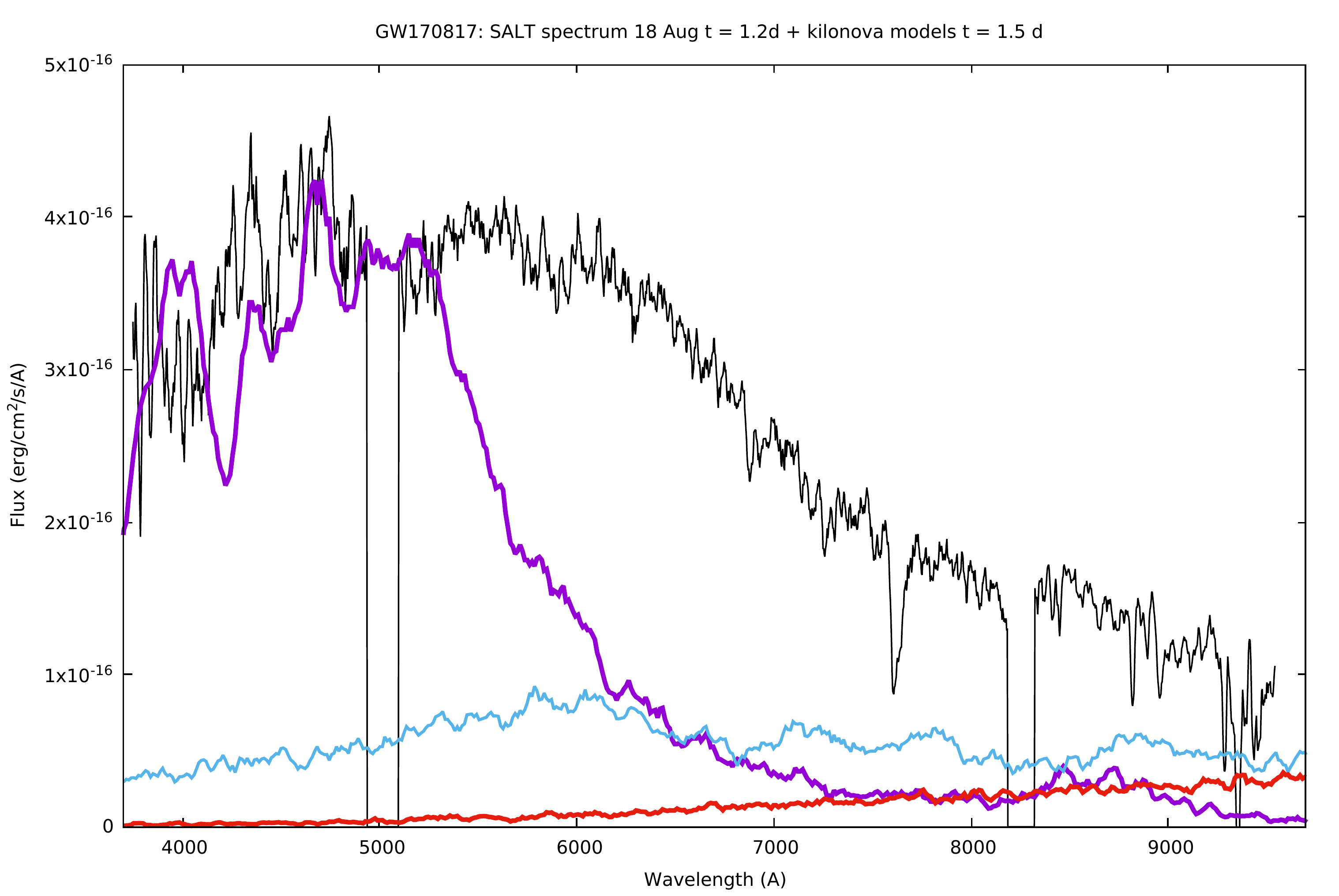}
    \caption{Comparison of the first SALT spectrum of AT 2017gfo, obtained 1.2 d after the GW event (in black), with two merger wind models of \citet{Tanaka2017}, namely $Y_e$ = 0.30 (purple) and $Y_e$ = 0.25 (blue), and for 1.5 d after a kilonova explosion, scaled to a distance of 40 Mpc. For comparison the higher velocity dynamical ejector model, APR4-1215, \citep{Tanaka2017}) is shown for comparison (red curve).The gaps in the SALT spectra at $\sim$5000\AA~and 8200\AA~are due to CCD gaps.}
    \label{fig:spec-1}
\end{figure}

\begin{figure}
	\includegraphics[width=\columnwidth]{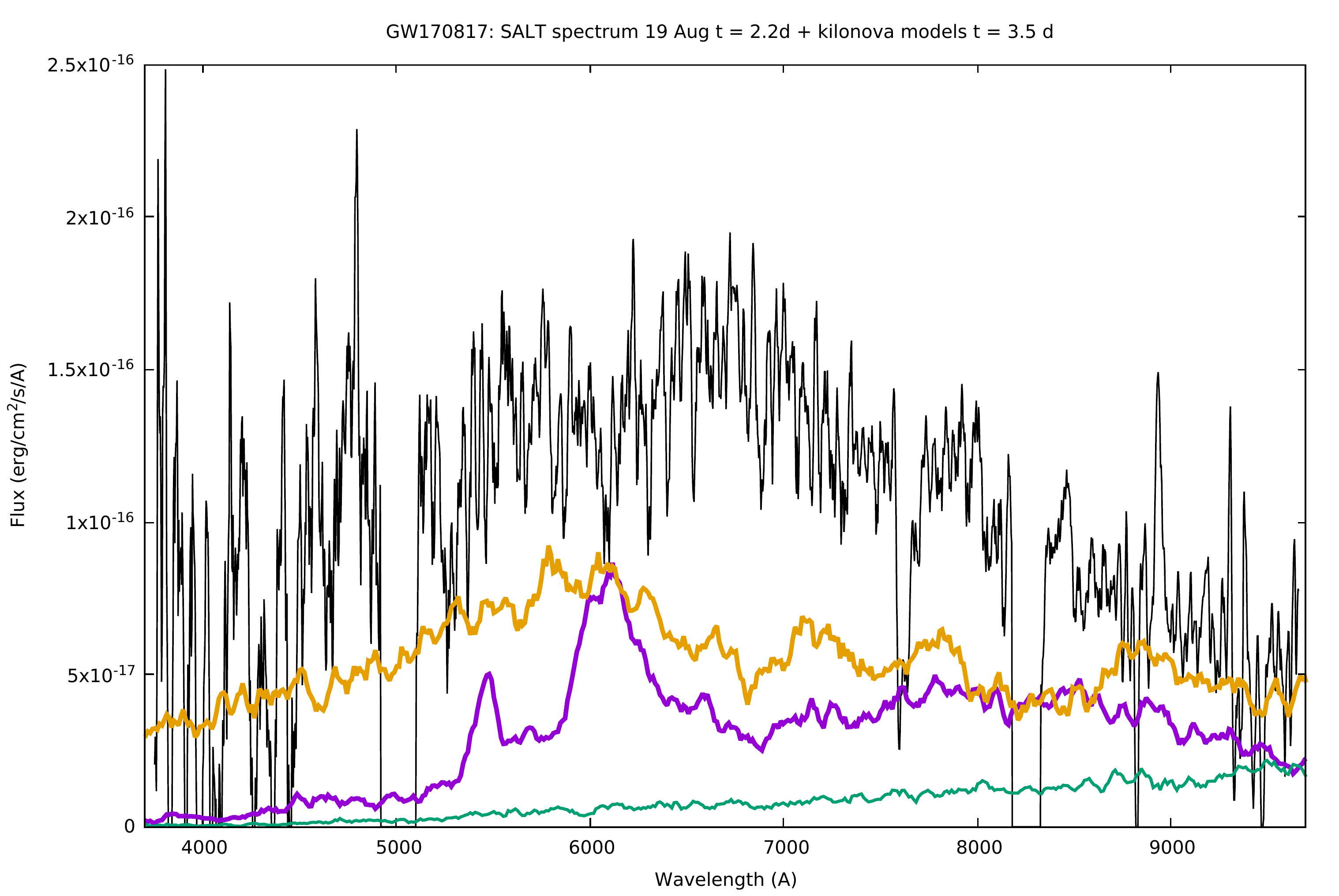}
    \caption{Similar plot to Fig. 1, comparing the second SALT spectrum of AT 2017gfo, obtained 2.2 d after the GW event (in black), with two wind models of \citet{Tanaka2017}, namely $Y_e$ = 0.30 (purple) and $Y_e$ = 0.25 (blue), for 3.5 d after a kilonova explosion, scaled to a distance of 40 Mpc. In addition we show the $Y_e$ = 0.25, t = 1.5 d model (orange), which is closer in delay time to the observation.}
    \label{fig:spec-2}
\end{figure}

\begin{figure}
	\includegraphics[width=\columnwidth]{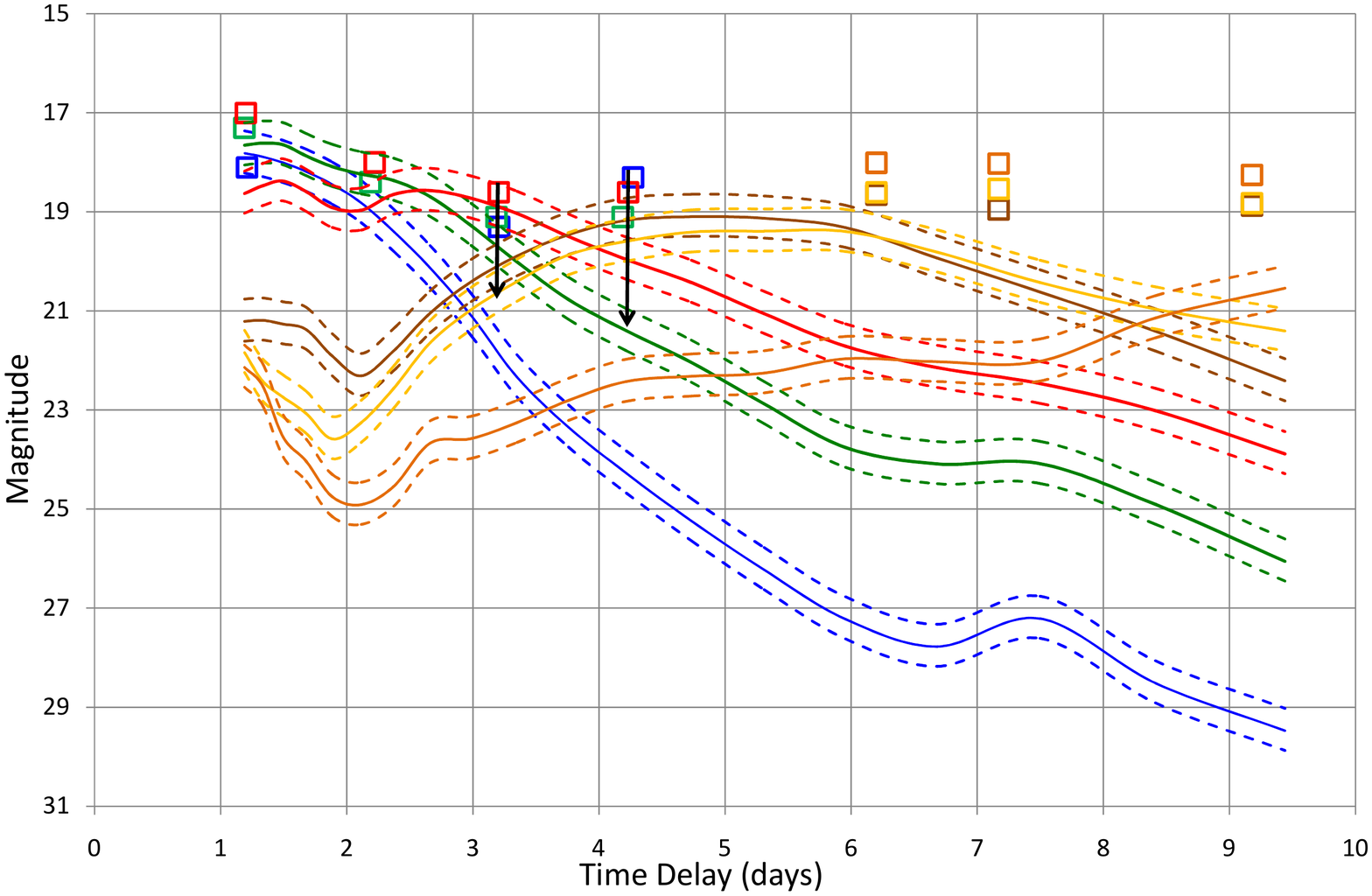}
    \caption{Comparison of optical/IR photometry of AT 2017gfo, obtained during the first $\sim$ 10 d after the GW event, with the $Y_e$ = 0.25 kilonova wind model of \citet{Tanaka2017}. The solid lines are the predicted magnitudes based on a distance of d = 40 Mpc, while the dashed lines represent the $\pm$ 8 Mpc distance uncertainty. The observed magnitudes and models are colour-coded for B (blue), V/W (green), R (red), JHK (yellow/brown/orange), while the arrows indicate brightness upper limits for $BVR$ measurements made after t = 3 d.  }
    \label{fig:phot-1}
\end{figure}

\begin{figure}
	\includegraphics[width=\columnwidth]{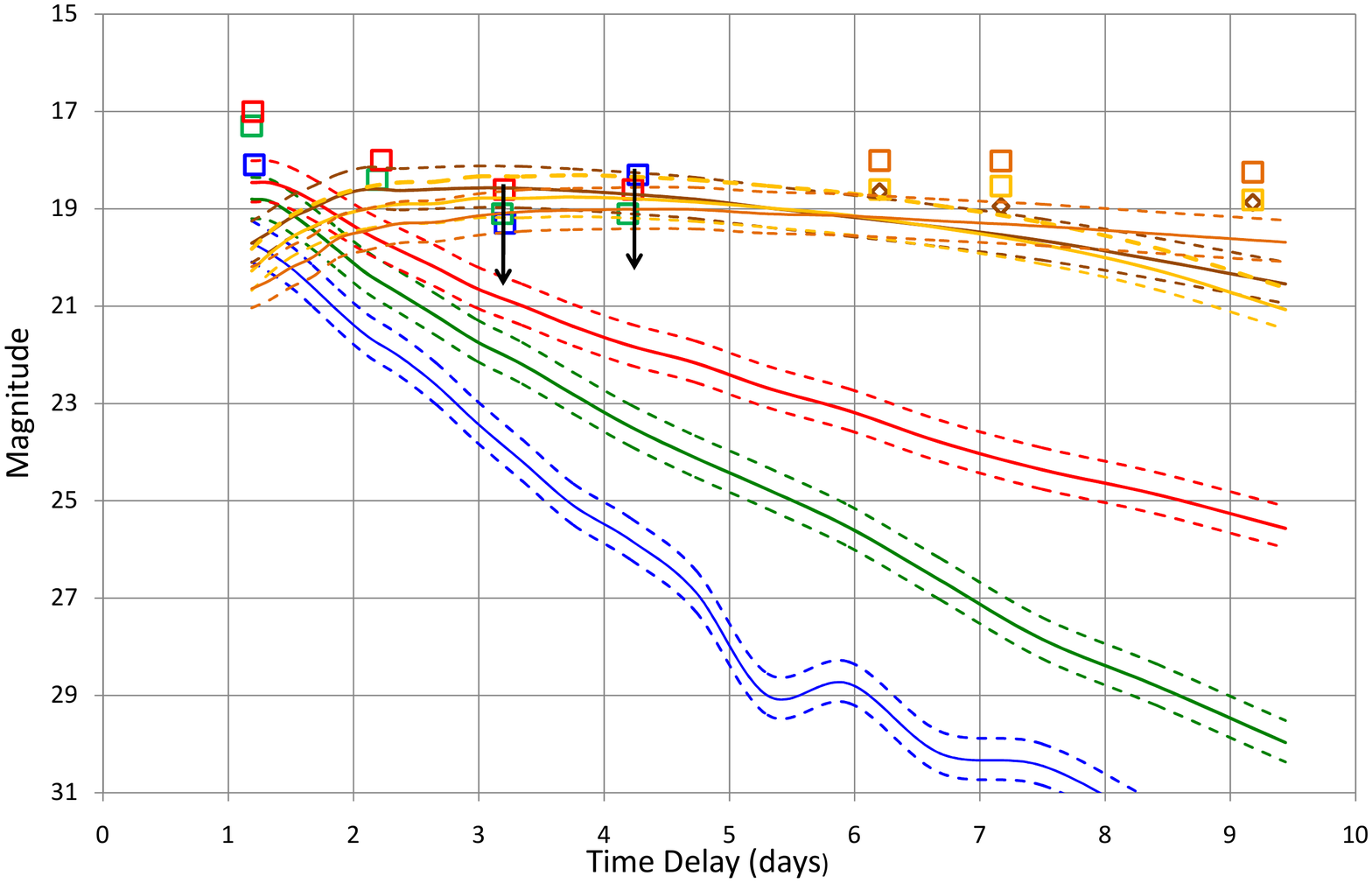}
    \caption{Same as Fig. 3, but compared to the $Y_e$ = 0.30 kilonova wind model of \citet{Tanaka2017}.  
    }
    \label{fig:phot-2}
\end{figure}

\begin{table*}

\centering
	\caption{Observing Details}
	\label{tab:table1}
	\begin{tabular}{llccccclc} 
		\hline
		Date & Start Time & Obs Type & Telescope & Filter/Bandpass & Exp Time & Mag/error & Conditions & Delay\\
        & (UTC) & & & & (s) & & & (d)\\
		\hline
18 Aug & 17:07:20 &	Spec & SALT & 3750$-$9600\AA	& 433 &	- & Cirrus; 1.2'' & 1.19\\
18 Aug & 17:06:55 &	Phot & MASTER-SAAO & W & 1080 & 17.3 $\pm$ 0.2 & Cirrus; 1.2'' & 1.19\\
18 Aug & 17:17:33 &	Phot & MASTER-SAAO & R & 540 & 17.0 $\pm$ 0.2 & Cirrus; 1.2'' & 1.20\\
18 Aug & 17:34:04 &	Phot & MASTER-SAAO & B & 540 & 18.1 $\pm$ 0.1 & Cirrus; 1.2'' & 1.21\\
19 Aug & 16:58:32 &	Spec & SALT	& 3750$-$9600\AA	& 716 &	- &	Clear; 1.2'' & 2.18\\
19 Aug & 17:06:57 &	Phot & MASTER-SAAO & W & 1080 & 18.4 $\pm$ 0.2 & Clear; 1.2'' & 2.19\\
19 Aug & 17:53:34 &	Phot & MASTER-SAAO & R & 540 & 18.0 $\pm$ 0.3 & Clear; 1.2'' & 2.22\\
20 Aug & 17:04:36 &	Phot & MASTER-SAAO & W & 540 & $>$19.1 & Cirrus; 1.1'' & 3.19\\
20 Aug & 17:25:56 &	Phot & MASTER-SAAO & R & 540 & $>$18.6 & Cirrus; 1.1'' & 3.20\\
20 Aug & 17:36:32 &	Phot & MASTER-SAAO & B & 540 & $>$19.3 & Cirrus; 1.1'' & 3.21\\
21 Aug & 17:08:14 &	Phot & MASTER-SAAO & W & 540 & $>$19.1 & Cirrus; 1.5'' & 4.19\\
21 Aug & 18:06:12 &	Phot & MASTER-SAAO & R & 540 & $>$18.6 & Cirrus; 1.5'' & 4.23\\
21 Aug & 19:20:23 &	Phot & MASTER-SAAO & B & 540 & $>$18.3 & Cirrus; 1.5'' & 4.27\\
23 Aug & 17:22 & Phot & IRSF & J & 1800 & 18.65/0.19 & Clear; 1.5'' & 6.20\\
23 Aug & 17:22 & Phot & IRSF & H & 1800	& 18.60/0.18 & Clear; 1.5'' & 6.20\\
23 Aug & 17:22 & Phot & IRSF & K & 1800	& 18.01/0.10 & Clear; 1.5'' & 6.20\\
24 Aug & 16:51 & Phot & IRSF & J & 2400	& 18.95/0.32 & Clear; N/A & 7.17\\
24 Aug & 16:51 & Phot & IRSF & H & 2400	& 18.53/0.17 & Clear; N/A & 7.17\\
24 Aug & 16:51 & Phot & IRSF & K & 2400	& 18.02/0.12 & Clear; N/A & 7.17\\
26 Aug & 16:57 & Phot & IRSF & J & 3000	& 18.87/0.30 & Clear; 1.3'' & 9.18\\
26 Aug & 16:57 & Phot & IRSF & H & 3000	& 18.82/0.23 & Clear; 1.3'' & 9.18\\
26 Aug & 16:57 & Phot & IRSF & K & 3000	& 18.25/0.25 & Clear; 1.3''	& 9.18\\
		\hline
	\end{tabular}
\end{table*}

\section{Conclusions}
We have presented optical and infrared observations from SALT and SAAO of the first optical counterpart (AT 2017gfo) of a gravitational wave source, GW170817, a kilonova explosion resulting from the merger of two neutron stars. 

SALT was the third telescope to observe AT 2017gfo \citep{Abbott2017, Andreoni2017}. Our early-time (1.2 $-$ 2.2 d) SALT spectra shows a relatively blue object, which is broadly consistent with the post-merger kilonova ejection models of \citet{Tanaka2017}.  The relatively blue colours are also consistent with the lower opacity of the Lathanide-free r-process elements in the ejector, although the expected features due to r-process elements are not seen. In comparing our spectroscopic and photometric measurements to the kilonova models of \citet{Tanaka2017}, we have concluded that there is qualitative agreement with the models invoking post-merger ejection of material out of the orbital plane. However, neither of these models match the observed spectra in their entirety. While the $Y_e$ = 0.30 model seems to better match the initial spectral shape and energetics, at least in the blue, the photometric evolution is closer the the $Y_e$ = 0.25 model predictions, notwithstanding that the fluxes are too low for the assumed distance of 40 Mpc. Recently \citet{Tanaka2017b} also concluded that the $Y_e$ = 0.25 model was a better match to photometry they reported of AT 2017gfo, which extended to t = 15 d after merger.    

These models predict an initial blue spectral energy distribution followed by strong wavelength-dependent dimming after the kilonova explosion, which are consistent with our optical/IR photometric observations. In particular, while at optical wavelengths ($BVR$) there is a significant dimming over a timescale of $\sim$ 2$-$3 d, the $JHK$ fluxes remained fairly constant, at least up to $\sim$ 9 d following the kilonova eruption.

The detection of an electromagnetic counterpart to a gravitational wave source, coming only $\sim$2 years after the first confirmed detection of such a source, bodes well for the study of future GW neutron star merger events. The ability of SALT to respond promptly and appropriately to transient alerts, in this case the GW170817 event, is one reason for the success of the observations reported here and will hopefully result in similar successes in the future. 

\section*{Acknowledgements}
We thank Professor Masaomi Tanaka for making available a recent preprint and model predictions for both spectra and photometric magnitudes of kilonovae. 

Some of the observations reported in this paper were obtained with the Southern African Large Telescope (SALT) under the Director's Discretionary Time programme 2017-1-DDT-009.

This research was financially supported for some of us (DAHB, SB, SMC, SBP, ERC, PV and TBB) by the National Research Foundation (NRF) of South Africa. MG acknowledges the Polish NCN grant OPUS 2015/17/B/ST9/03167. J Mao is supported by the Hundred Talent Program, the Major Program of the Chinese Academy ofSciences (KJZD-EW-M06), the National Natural Science Foundation of China 11673062, and the Oversea Talent Program of Yunnan Province. Part of this research was funded by the Australian Research Council Centre of Excellence for Gravitational Wave Discovery (OzGrav), CE170100004.  JC acknowledges the Australian Research Council Future fellowship grant FT130101219.

\bsp	
\label{lastpage}
\end{document}